\documentclass[nofootinbib,preprint,superscriptaddress,showpacs,amsmath,amssymb,aps,pra,showkeys]{revtex4-2}

\usepackage{amsmath,amsfonts,amssymb}
\usepackage{graphicx}
\usepackage{xcolor}
\usepackage{float}

\usepackage{calc}

\usepackage{accents}

\DeclareMathOperator{\arccosh}{arcosh}

\begin{document}

\title{Multielectron effect in strong-field ionization of CO}

\author{Mahmoud Abu-samha}
\affiliation{College of Engineering and Technology, American University of the Middle East, Kuwait}

\author{L. B. Madsen}
\affiliation{Department of Physics and Astronomy, Aarhus University, 8000 \r{A}arhus C, Denmark}

\author{N. I. Shvetsov-Shilovski}
\email{n79@narod.ru}
\affiliation{Institut f\"{u}r Theoretische Physik, Leibniz Universit\"{a}t Hannover, 30167 Hannover, Germany}

\date{\today}

\begin{abstract}
We investigate the effects of the multielectron polarization of the ion described by the induced dipole potential in photoelectron momentum distributions produced in ionization of the CO molecule by a strong laser field. We present results of the numerical solution of the time-dependent Schr\"{o}dinger equation in three spatial dimensions and semiclassical simulations accounting for quantum interference. We predict the change of the asymmetry and interference patterns in two-dimensional photoelectron momentum distributions as well as longitudinal momentum distributions. 
By using a semiclassical model we identify the mechanism responsible for the observed effects. It is shown that the modifications of electron momentum distributions are caused by a combined effect of the force acting on photoelectrons due to the induced dipole potential and the linear Stark-shift of the ionization potential.  
\end{abstract}



\maketitle

\newpage

\section{Introduction} 
Interaction of strong laser pulses with atoms and molecules leads to such highly nonlinear phenomena as above-threshold ionization (ATI), formation of the high-energy plateau in the electron spectrum, generation of high-order harmonics (HHG), and nonsequential double ionization (NSDI) (see Refs.~\cite{Graefe2016,Lin2018} for recent reviews). Among the main theoretical methods used in strong-field physics are the strong-field approximation (SFA) \cite{Keldysh1964,Faisal1973,Reiss1980}, the direct numerical solution of the time-dependent Schr\"{o}dinger equation (TDSE) (see, e.g., Refs. \cite{Muller1999,Bauer2006,Madsen2007,Patchkovskii2017} and references therein), and the semiclassical models applying classical mechanics to describe the electron motion in the continuum (e.g., the two-step \cite{Linden88,Gallagher88,Corkum89} and the three-step \cite{Kulander_Schafer,Corkum1993} models). 

All these approaches usually employ the single-active-electron (SAE) approximation \cite{Kulander1988,Kulander1991}. In the SAE a many-electron atom or molecule ionized by a strong laser pulse is considered as a one-electron system. The single active electron moves in the laser field and an effective potential, which reproduces the ground state and singly excited states. The SAE approximation provides a reliable basis for understanding of ATI and HHG \cite{BeckerRev2002,GrossmannBook}. However, in recent years, multielectron effects have been attracting considerable attention (see, e.g., Refs.~\cite{Kang2018,Le2018,Shvetsov2018,Mahmoud2020,Mahmoud2022} and references therein).

The multielectron polarization (MEP) effect in ATI, i.e., polarization of the ionic core in a laser field, has been intensively studied for different atoms and molecules \cite{Dimitrovski2010,Pfeiffer2012,Shvetsov2012,Dimitrovski2014,Dimitrovski2015,Dimitrovski2015JPB,Majety2017,Shvetsov2018,Mahmoud2020,Mahmoud2022}. The CO molecule is one of the well-known examples where account of the MEP effect is necessary. Significant progress has been achieved in studies of the MEP effects in strong-field ionization and generation of high-order harmonic from this molecule, see. e.g., Refs.~\cite{Zhang2013,Hoang2017,Ohmura2018,Song2017,Le2018}. It was shown that the MEP effect strongly affects the orientation dependence of the ionization yield in CO \cite{Zhang2013}. In Ref.~\cite{Hoang2017} laser-induced polarization of the ionic core was accounted within the SAE approximation. The study of Ref.~\cite{Hoang2017}, as well as many other studies of MEP effects, is based on the effective potential for the single outer electron, which is derived in the adiabatic approximation in Refs.~\cite{Brabec2005,Zhao2007} and \cite{Dimitrovski2010}. This potential takes into account the laser field, the Coulomb potential, and the polarization effects of the ionic core [see Eq.~(\ref{pot}) in Sec. II B].  

It should be noted that the solution of the TDSE within the SAE approximation causes for the CO molecule the following problem: instead of ionization, the external field may drive the active electron to lower-lying orbitals. Simultaneously, dipole coupling to these lower-lying bound states of the potential strongly depends on molecular orientation. As a result, the dipole transitions and shifting of population from the highest occupied molecular orbital (HOMO) to the lower bound states significantly affect the total ionization yield at different orientation angles. A solution of this problem was found in Ref.~\cite{Mahmoud2020}. It was shown that by including MEP, the external field is turned off within the molecular radius, and therefore the dipole coupling becomes negligible. Furthermore, as long as the laser field is turned off within the molecular radius, the long-range induced dipole term in the MEP potential is not required for a correct description of the orientation dependence of the total ionization yield \cite{Mahmoud2020}. The study of Ref.~\cite{Mahmoud2020} considered not only total ionization yields, but also photoelectron momentum distributions (PMDs). No pronounced MEP effects were found in PMDs produced by short ($2$ optical cycles) laser pulses with the intensity of $8.8\times10^{13}$ W/cm$^2$. However, longer and more intense laser pulses were not considered in Ref.~\cite{Mahmoud2020}. 

In this paper we investigate the MEP effects in momentum distributions produced in strong-field ionization of the CO molecule. As in Refs.~\cite{Hoang2017,Mahmoud2020}, we solve the TDSE within the SAE approximation and use the potential following from the treatment of Refs.~\cite{Brabec2005,Zhao2007,Dimitrovski2010}. We find a pronounced MEP effect in two-dimensional electron momentum distributions as well as in longitudinal momentum distributions along the polarization direction. In order to reveal the mechanism underlying this effect, we apply a semiclassical model that allows us to understand how the MEP affects relevant electron trajectories.  

It was for the first time shown in Ref.~\cite{Pfeiffer2012} that the TDSE with the effective potential \cite{Brabec2005,Zhao2007,Dimitrovski2010} and accounting for the Stark shift of the ionization potential can be approximately separated in the parabolic coordinates. The separation procedure leads to a certain tunneling geometry, and the resulting physical picture is referred to as tunnel ionization in parabolic coordinates with induced dipole and Stark shift (TIPIS). The semiclassical model applying the TIPIS approach has been widely used in studies of MEP effects in circularly or close to circularly polarized laser fields, see, e.g., Refs.~\cite{Pfeiffer2012,Shvetsov2012,Dimitrovski2014,Dimitrovski2015,Dimitrovski2015JPB}. Predictions of this model that does not describe quantum interference are shown to be in a good agreement with experimental results \cite{Pfeiffer2012,Dimitrovski2014,Dimitrovski2015JPB} and TDSE calculations (see Refs.~\cite{Pfeiffer2012,Shvetsov2012}).

It should be noted that the effective potential of Refs.~\cite{Brabec2005,Zhao2007,Dimitrovski2010} is applicable at large and intermediate distances, but not in the vicinity of the ionic core. It is known that the majority of electrons generated in strong-field ionization do not return to their parent ions. These electrons are known as direct electrons. There are also electrons that are driven back by the oscillating laser field to their ions and scatter from them by large angles. These rescattered electrons form the high-energy plateau in the electron spectrum. Since the rescattering process is suppressed in circularly and close to circularly polarized laser fields, the vast majority of electron trajectories in such fields do not return to their parent ions.  

However, the situation is different in linearly polarized field, and the applicability of the semiclassical model based on the TIPIS approach for linear polarization raised questions. This problem was addressed in Ref.~\cite{Shvetsov2018} that combines the TIPIS approach with the semiclassical two-step model (SCTS) for strong-field ionization \cite{Shvetsov2016}. A simple procedure that allows to identify the domain of a PMD that can be reliably treated by the TIPIS approach was proposed in Ref.~\cite{Shvetsov2018}. The resulting semiclassical method describes quantum interference by accounting for the effective potential beyond the semiclassical perturbation theory. It was shown that the presence of the MEP term in the effective potential leads to a narrowing of the longitudinal momentum distributions due to electron focusing by the induced dipole potential and modification of interference structures \cite{Shvetsov2018}. Here we further develop and modify the model of Ref.~\cite{Shvetsov2018} in order to make it applicable to ionization of the CO molecule.

The paper is organized as follows. In Sec.~II we discuss the approach we use to solve the TDSE and the semiclassical model for the CO molecule. In Sec.~III we present results of our simulations and discuss the MEP effect in electron momentum distributions. We analyze the mechanism responsible for the MEP effect. The conclusions are given in Sec.~IV. Atomic units are used throughout the paper unless indicated otherwise. 

\section{Computational details}

\subsection{TDSE method}

The ionic potential obtained in Refs.~\cite{Brabec2005,Zhao2007,Dimitrovski2010} at large distances reads as
\begin{equation}
V\left(\vec{r},t\right)=-\frac{Z}{r}-\frac{\left(\vec{\mu}_{\textrm{p}}+\vec{\mu}_{\textrm{ind}}\right)\cdot\vec{r}}{r^3},
\label{pot}
\end{equation}
where $\vec{\mu}_{\textrm{p}}$ and $\vec{\mu}_{\textrm{ind}}$ are the permanent and induced dipoles of the cation, respectively. The MEP term in Eq.~(\ref{pot}) can be written as $-\vec{\mu}_{\textrm{ind}}\cdot \vec{r}/r^3=-\left[\boldsymbol{\alpha}\cdot \vec{E}\right]\cdot \vec{r}/r^3$, where $\boldsymbol{\alpha}$ is the polarizability tensor and $\vec{E}\left(t\right)$ is the electric field. For a linear molecule ionized by a field linearly polarized along the $z$-axis, the product $\boldsymbol{\alpha}\cdot \vec{E}\left(t\right)$ reduces to $\left(\alpha_{\perp}E_{z},0,\alpha_{||}E_{z}\right)$, where $\alpha_{\perp}$ and $\alpha_{||}$ are the components of the polarizability tensor perpendicular and parallel to the laser polarization, respectively. These components are calculated from the molecular-fixed frame components $\alpha_{xx}$ and $\alpha_{zz}$:
\begin{align}
\alpha_{\perp}&=\left(\alpha_{zz}-\alpha_{xx}\right)\sin\left(\beta\right)\cos\left(\beta\right),\\
\alpha_{||}&=\alpha_{xx}\sin^2\left(\beta\right)+\alpha_{zz}\cos^2\left(\beta\right). 
\label{comp}
\end{align}
In what follows we consider only orientation angle $\beta=0^{\circ}$ , at which $\alpha_{\perp}$ and the corresponding ($x$) induced dipole component vanish. Therefore, the MEP term reads as $-\alpha_{||}\vec{E}\left(t\right)\cdot \vec{r}/r^{3}$. 

To include MEP in the TDSE calculations for CO, the polarizability $\alpha_{||}$ was calculated for the CO$^{+}$ cation at the frozen geometry of the neutral CO molecule (with a C–O bond distance of $2.13$~a.u.). The polarizability was determined from quantum chemistry calculations within the framework of density functional theory employing the local spin-density approximation (LSDA) \cite{Vosko1980} and Augcc-pVDZ basis set \cite{Dunning1989}. The CO$^{+}$ cation has a static polarizability tensor with non-zero components $\alpha_{xx}=7.88$~a.u. and $\alpha_{zz}=12.39$~a.u. The dynamic polarizability was computed for the CO$^{+}$ cation at an external field frequency ($\omega=0.057$~a.u.) corresponding to $800$-nm wavelength, and the produced polarizability tensor has the following non-zero components $\alpha_{xx}=\alpha_{yy}=8.22$~a.u. and $\alpha_{zz}=12.63$~a.u. Since the difference between the static and dynamic polarizabilities of the CO$^{+}$ cation is negligible, we use the static polarizabilities in our TDSE calculations. At $\beta=0^{\circ}$, the molecular and lab-frame $z$-axes are aligned, and in this case $\alpha_{||}=\alpha_{zz}=12.39$~a.u.

It is assumed that the MEP effect cancels the external field at $r<r_{c}$, where $r_{c}=\alpha_{||}^{1/3}$ is the cutoff radius \cite{Brabec2005,Zhao2007,Kang2018}. This cancellation can be seen as a result of the polarization of the remaining electrons, which rearrange until they no longer feel an effective field. In the length gauge, the interaction of the photoelectron with the laser field is described by the following term in the TDSE (neglecting the MEP):
\begin{equation}
V_{LG}^{\textrm{Ext}}\left(\vec{r},t\right)=E\left(t\right)r\cos\left(\theta\right)=E\left(t\right)\sqrt{\frac{2}{3}}r\bar{P}_{l}\left(\zeta\right).
\end{equation}
Here $\bar{P}_{l}$ is a normalized Legendre function, $\zeta=\cos \theta$, and $\theta$ is the polar angle. With consideration of the MEP effect as described with the potential of Eq.~(\ref{pot}), this term reads as
\begin{equation}
  V_{LG}^{\textrm{Ext}}\left(\vec{r},t\right)=
    \begin{cases}
      & \left(1-\frac{\alpha_{\par}}{r^3}\right)E\left(t\right)\sqrt{\frac{2}{3}}r\bar{P}_{l}\left(\zeta\right), r>r_{c}\\
      & 0, r\leq r_{c},\\
    \end{cases}       
\end{equation}
where the MEP term is zero for $r<r_{c}$. The effective potential used in the TDSE is 
\begin{equation}
V_{\textrm{eff}}\left(\vec{r},t\right)=V\left(\vec{r}\right)+V_{LG}^{\textrm{Ext}}\left(\vec{r},t\right),
\label{pot_tdse}
\end{equation}
where $V\left(\vec{r}\right)$ is the SAE potential of the CO molecule.  

The SAE potential describing CO was determined from quantum chemistry calculations following the procedure in Ref.~\cite{Mahmoud2010}. The molecule is placed along the molecular frame $z$-axis such that the center-of-mass coincides with the origin and the O atom points in the positive $z$ direction.
The SAE potential of CO was expanded in partial waves:
\begin{equation}
V\left(\vec{r}\right)=\sum_{l,m=0}^{l_{\textrm{max}}}V_{l0}\left(r\right)Y_{l0}\left(\theta,\phi\right),
\label{pot_sae}
\end{equation}
where $m=0$ since the molecule is linear. The expansion was truncated at $l_{\textrm{max}}=20$. Based on our SAE potential for CO, the HOMO of CO is the $5\sigma$ with energy $-0.542$~a.u. in reasonable agreement with the literature value of $-0.555$~a.u. \cite{Kobus1993}. The potential (\ref{pot_sae}) is obtained for $0<r<R_{0}$, where $R_0=320$~a.u. At $r=R_{0}$ our SAE potential coincides with the Coulomb potential $-Z/r$ with the effective charge $Z$ equal to $0.88$. The latter value was obtained from quantum chemistry calculations in a smaller box with the radius of $10.0$~a.u. Therefore, for $r>R_{0}$ the molecule is described by the Coulomb potential with $Z=0.88$.

We follow the split-operator spectral method of Hermann and Fleck \cite{Hermann1988} (see Ref.~\cite{Kjeldsen2007} for details of our implementation)
to obtain the wavefunction of the HOMO of the CO molecule. Starting with a guess initial wavefunction (Hydrogen 2$p_z$ state), denoted $\Psi\left(r,t=0\right)$ and the SAE potential for the CO molecule, we perform field-free propagation for $1000$~a.u. and save the wavepacket every $1.0$~a.u. From the time-dependent wavefunction, $\Psi\left(r,t\right)$, a bound-state spectrum $P\left(E\right)$ is produced with the aid of the autocorrelation function $P\left(t\right)$, as
\begin{equation}
\mathcal{P}\left(E\right)=\frac{1}{T}\int_{0}^{T}dt\omega\left(t\right)\exp\left(iEt\right)\mathcal{P}\left(t\right),
\label{auto}
\end{equation}
where $T$ is the field-free propagation time, $\omega\left(t\right)$ is the Hanning window function \cite{Hermann1988}, and $\mathcal{P}=\langle \Psi\left(r,t=0\right)|\Psi\left(r,t\right) \rangle$. Once the orbital energy of the HOMO, $E_{\textrm{HOMO}}$, is well resolved in the $\mathcal{P}\left(E\right)$ spectrum, the corresponding wavefunction can be constructed and normalized as follows
\begin{equation}
\Psi_{\textrm{HOMO}}\left(r\right)=\frac{1}{T}\int_{0}^{T}dt\omega\left(t\right)\exp\left(iE_{\textrm{HOMO}}t\right)\Psi\left(r,t\right).
\label{psihomo}
\end{equation}

The electric field $\vec{E}\left(t\right)$, linearly-polarized along the laboratory-frame $z$-axis, is defined as
\begin{equation}
\vec{E}=-\partial_{t}A\left(t\right)\vec{e}_{z}=-\partial_{t}\left(\frac{E_0}{\omega}\sin^2\left(\pi t / \tau\right)\cos\left(\omega t+\phi\right)\right)\vec{e}_{z},
\label{laser_field}
\end{equation}
where $E_0$ is the field amplitude, $\omega$ is the frequency, $\phi$ is the carrier-envelope phase (CEP) for a laser pulse with duration $\tau$, and $\vec{e}_{z}$ is a unit vector. The TDSE calculations were performed at laser frequency $\omega=0.057$~a.u. corresponding to $800$~nm wavelength and peak electric field values of $E_0= 0.05$ and $0.1$~a.u., corresponding to laser intensities of $8.8\times10^{13}$ and $3.51\times10^{14}$~W/cm$^2$. The CEP value is $\phi= -\pi/2$, and the
pulse lengths are specified below.

In the TDSE calculations, the radial grid contains $4096$ points and extends to $400$~a.u. The size of the angular basis set is limited by $l_{\textrm{max}}=60$. The calculations were performed at orientation angle $\beta=0^{\circ}$. The PMDs were produced by projecting the wavepacket at the end of the
laser pulse on Coulomb scattering states in the asymptotic region ($r>20$~a.u.), an approach that was validated in Ref.~\cite{Madsen2007} and recently applied for molecular hydrogen \cite{Mahmoud2016,Mahmoud2017,Mahmoud2022}.

\subsection{Semiclassical model}

In semiclassical models the electron trajectory is found from integrating of Newton's equation of motion:
\begin{equation}
\frac{d^2\vec{r}}{dt^2}=-\vec{\nabla}V_{\textrm{eff}}\left(\vec{r},t\right)
\label{newton}
\end{equation}
where the effective potential $V_{\textrm{eff}}\left(\vec{r},t\right)$ is given by Eq.~(\ref{pot_tdse}). In order to solve Eq.~(\ref{newton}), the initial conditions, i.e., initial electron velocity and the starting point of the trajectory (the tunnel exit point), are needed. In the TIPIS model, the tunnel exit is found as $z_{e}\approx-\eta_{e}/2$, where $\eta_{e}$ satisfies the following equation (see Ref.~\cite{Pfeiffer2012}):
\begin{equation}
\frac{\beta_{2}\left(E\right)}{2\eta}+\frac{m^2-1}{8\eta^2}-\frac{E\eta}{8}+\frac{\alpha_{||}E}{\eta^2}=-\frac{I_{p}\left(E\right)}{4},
\label{tipis_exit}
\end{equation}
where $I_{p}\left(E\right)$ is the Stark-shifted ionization potential, $m$ is the magnetic quantum number of the initial state, and
\begin{equation}
\beta_2\left(E\right)=Z-\left(1+\left|m\right|\right)\frac{\sqrt{2I_{p}\left(E\right)}}{2}. 
\end{equation}
The ionization potential $I_{p}\left(E\right)$ is given by
\begin{equation}
I_{p}\left(E\right)=I_{p}\left(0\right)+\vec{\mu}_{\textrm{\scriptsize{HOMO}}}\cdot{\vec{E}}+\frac{1}{2}\alpha_{||,\textrm{\scriptsize{HOMO}}}\vec{E}^2.
\label{ion_pot}
\end{equation}
Here $I_{p}\left(0\right)$ is the ionization potential when the field is absent, $\vec{\mu}_{\textrm{\scriptsize{HOMO}}}=\vec{\mu}_{\textrm{N}}-\vec{\mu}_{\textrm{I}}$ and $\alpha_{||,\textrm{\scriptsize{HOMO}}}=\alpha_{||,N}-\alpha_{||,I}$ are the dipole moment and static polarizability of the HOMO, respectively. Here, in turn, the index $N$ refers to the neutral molecule, and the index $I$ corresponds to its ion. The instantaneous value of the laser field $E\left(t_0\right)$ at ionization time $t_0$ should be used instead of the static field $E$ in Eqs.~(\ref{tipis_exit})-(\ref{ion_pot}). 

It is assumed in the SCTS model that the electron starts with zero initial velocity along the laser polarization direction: $v_z=0$. However, it can have a nonzero initial transverse velocity $\vec{v}_{0,\perp}$. Along with the ionization time $t_0$, this initial transverse velocity completely determines the electron trajectory. In the SCTS model, $t_{0}$ and $v_{0,\perp}$ are distributed in accord with the static ionization rate \cite{Delone1991}:
\begin{equation}
w\left(t_0,v_{0,\perp}\right)\sim\exp\left[-\frac{2\kappa^3}{3E\left(t_0\right)}\right]\exp\left[-\frac{\kappa v_{0,\perp}^2}{E\left(t_0\right)}\right],
\label{tun_rate}
\end{equation}
where $\kappa=\sqrt{2I_{p}\left(E\right)}$. 

However, the semiclassical model described here does not agree very well with TDSE results for the CO molecule, especially at intensities of the order of $3-4\times10^{14}$ W/cm$^2$. The reason for this is that at such laser intensities the tunnel exit point predicted by the TIPIS model is relatively close to the origin, where the potential of Eq.~(\ref{pot}) is not valid. Indeed, for $E_{0}=0.1$ a.u., what corresponds to the laser intensity of $3.51\times10^{14}$ W/cm$^2$, the tunnel exit point calculated from Eq.~(\ref{ion_pot}) is $z_{e}=-6.5$~a.u. The inapplicability of the potential of Eq.~(\ref{pot}) at such distances results in a too strong asymmetry of PMDs (see Sec.~3).  

Therefore, we modify the semiclassical model as compared to the approach used in Ref.~\cite{Shvetsov2018}. Instead of Eq.~(\ref{tipis_exit}), we use the expression for the tunnel exit point predicted by the SFA, see, e.g., Refs.~\cite{Popruzhenko2008a,Popruzhenko2008b,Yan2010,Li2016,Brennecke2020}.  
For a sufficiently long laser pulse, for which the pulse envelope $E_0\sin^2\left(\pi t / \tau\right)$ is nearly constant during one half of the laser cycle, the exit point can be approximated as
\begin{equation}
z_{e}=\frac{E_{0}}{\omega^2}\sin^2\left(\frac{\pi t}{\tau}\right)\sin\left(\omega t+\phi\right)\left[1-\sqrt{\gamma\left(t_0,v_{0,\perp}\right)^2+1}\right],
\label{sfa_exit}
\end{equation}
where 
\begin{equation}
\gamma\left(t_0,v_{0,\perp}\right)=\frac{\omega \sqrt{2I_{p}+v_{0,\perp}^{2}}}{E_{0}\sin^2\left(\frac{\pi t}{\tau}\right)\sin\left(\omega t+\phi\right)}
\label{gamma_perp}
\end{equation}
is the effective Keldysh parameter (see Ref.~\cite{Li2016}). In accord with the SFA, a photoelectron has also the nonzero initial parallel velocity. Within the same assumption regarding the pulse envelope, the initial parallel velocity is given by
\begin{equation}
v_{0,z}=-\frac{E_{0}}{\omega}\sin^2\left(\frac{\pi t}{\tau}\right)\cos\left(\omega t+\phi\right)\left[\sqrt{\gamma\left(t_0,v_{0,\perp}\right)^2+1}-1\right].
\label{v0_par}
\end{equation}
We note that in the tunneling limit $\gamma\left(t_{0},v_{0,\perp}\right)\to 0$, and the initial longitudinal velocity vanishes.   

The semiclassical model based on the SFA expressions (\ref{sfa_exit}) and (\ref{v0_par}) is not a quasistatic approach. Therefore, the combination of this model with the quasistatic ionization rate (\ref{tun_rate}) is questionable. For this reason, we distribute $t_0$ and $v_{0,\perp}$ in accord to the instantaneous ionization rate \cite{Bondar2008}:
\begin{equation}
w\left(t_0,v_{0,\perp}\right)\propto\exp\left[-\frac{2I_p}{\omega}f\left(t_0,v_{||},v_{0,\perp}\right)\right],
\label{bondar_rate}
\end{equation}
where
\begin{equation}
f\left(t_0,v_{||},v_{0,\perp}\right)=\left(1+\frac{1}{2\gamma^2}+\frac{v^2}{2I_p}\right)\arccosh \alpha-\sqrt{\alpha^2-1}\left(\frac{\beta}{\gamma}\sqrt{\frac{2}{I_p}}v_{||}+\frac{\alpha\left[1-2\beta^2\right]}{2\gamma^2}\right).
\label{bond_f}
\end{equation}
Here, in turn, 
\begin{equation}
    \left\{
        \begin{array}{c}
                \alpha \\
                \beta
    	\end{array}
    \right\}=\frac{\gamma}{2}\left(\sqrt{\frac{v^2}{2I_p}+\frac{2}{\gamma}\frac{v_{||}}{\sqrt{2I_p}}+\frac{1}{\gamma^2}+1}\pm \sqrt{\frac{v^2}{2I_p}-\frac{2}{\gamma}\frac{v_{||}}{\sqrt{2I_p}}+\frac{1}{\gamma^2}+1}~\right),
		\label{bond_alpha}
\end{equation}
\begin{equation}
\gamma=\frac{\omega\sqrt{2I_p}}{E_0\sin^2\left(\pi t / \tau\right)},
\label{gamma_t}
\end{equation}
\begin{equation}
v_{||}=v_{0,z}-\frac{E_0}{\omega}\cos\left(\omega t+\phi\right),
\end{equation}
and
\begin{equation}
v^2=v_{||}^2+v_{0,\perp}^2.
\end{equation}
The resulting semiclassical approach shows good qualitative agreement with the TDSE results provided the Stark shift of the ionization potential is neglected in Eqs.~(\ref{bondar_rate})-(\ref{gamma_t}), i.e., $I_p=I_p\left(0\right)$, a feature we ascribe to the approach being non-quasistatic. 

With the initial conditions sampled according to these SFA formulas, Eq.~(\ref{newton}) is solved by using a fourth-order Runge-Kutta method with adaptive step size see, e.g., Ref.~\cite{Press2007}. Let us introduce $t_{1}=\max\left(\tau,t_{\textrm{out}}\right)$, where $t_{\textrm{out}}$ is the time, at which an electron leaves the circle of radius $R_{0}$ centered at the origin and $\tau$ is the end of the laser pulse. At $t\geq t_{1}$ the motion occurs only in the Coulomb field with the effective charge $Z=0.88$. We calculate the conserving energy $E$ and exclude the trajectories with $E<0$ as they contribute to population of the Rydberg states \cite{Eichmann2008,Shvetsov2009}. The electron momentum $\vec{p}\left(t_1\right)$ and position $\vec{r}\left(t_1\right)$ determine the asymptotic (final) momentum $\vec{k}$. Indeed, the magnitude of the final momentum can be found from the energy conservation
\begin{equation}
\frac{k^2}{2}=\frac{p^2\left(t_1\right)}{2}-\frac{Z}{r\left(t_1\right)},
\label{enr_cons}
\end{equation}
and the direction of $k$ is determined by the following expression (see Refs.~\cite{Shvetsov2009,Shvetsov2012}):   
\begin{equation}
\vec{k}=k\frac{k(\vec{L}\times\vec{a})-\vec{a}}{1+k^2L^2},
\label{gor_form}
\end{equation}
where $\vec{L}=\vec{r}\left(t_1\right)\times \vec{p}\left(t_1\right)$ and $\vec{a}=\vec{p}\left(t_1\right)\times \vec{L}-\vec{r}\left(t_1\right)/r\left(t_1\right)$ are the angular momentum and the Runge-Lenz vector, respectively.  

The system consisting of the linearly polarized laser field and the CO molecule oriented along the polarization direction possesses cylindrical symmetry.
As in Refs.~\cite{Shvetsov2016,Shvetsov2018,Brennecke2020} we perform the simulation in two dimensions (2D) and obtain the results for the three-dimensional (3D) system. To this end, we use the simple relation between the Jacobian $J_{2D}$ in the 2D case and the 3D Jacobean $J_{3D}$ for a cylindrically symmetric system derived in Ref.~\cite{Brennecke2020}: 
\begin{equation}
\left|J_{3D}\right|=\frac{k_{\perp}}{v_{0,\perp}}\left|J_{2D}\right|,
\label{jac}
\end{equation}
where $k_{\perp}=\sqrt{k_{x}^2+k_{y}^2}$. This relation, which has been already used in Ref.~\cite{Shvetsov2016}, allows us to reduce the computational costs of semiclassical model significantly. 

As in Ref.~\cite{Shvetsov2018}, we assign the SCTS phase to every trajectory. This phase reads as (see Ref.~\cite{Shvetsov2016}):
\begin{equation}
\Phi\left(t_0,\vec{v}_{0}\right)=\vec{v_0}\cdot \vec{r}\left(t_0\right)+I_{p}t_{0}-\int_{t_0}^{\infty}dt\left\{\frac{p^2\left(t\right)}{2}+V_{\textrm{ion}}\left[\vec{r}\left(t\right)\right]-\vec{r}\left(t\right)\cdot \vec{\nabla}V_{\textrm{ion}}\left[\vec{r}\left(t\right)\right]\right\}.
\end{equation}
Here we introduce the ionic potential $V_{\textrm{ion}}\left(\vec{r},t\right)=V_{\textrm{eff}}\left(\vec{r},t\right)-\vec{E}\left(t\right)\cdot\vec{r}$. 

Since we deal with two different potentials, namely the SAE potential of Eq.~(\ref{pot_sae}) describing the CO molecule for $r<R_{0}$ and the Coulomb potential $V_{0}=Z/r$ for $r>R_{0}$, we need to distinguish between two different cases. In the first case, where an electron leaves the circle of radius $R_0$ at the time $t_{\textrm{out}}$ after the end of the laser pulse at $t=\tau$, the SCTS phase reads as
\begin{align}
\Phi\left(t_0,\vec{v}_{0}\right)&=\int_{t_0}^{\tau}dt\left\{\frac{p^2\left(t\right)}{2}+V_{\textrm{ion}}\left[\vec{r}\left(t\right)\right]-\vec{r}\left(t\right)\cdot \vec{\nabla}V_{\textrm{ion}}\left[\vec{r}\left(t\right)\right]\right\}+E\left(t_{\textrm{out}}-\tau\right) \nonumber \\
&+\int_{\tau}^{t_{\textrm{out}}}dt\left\{-\vec{r}\cdot \vec{\nabla}V_{\textrm{ion}}\left[\vec{r}\left(t\right)\right]\right\}+\lim_{t\to\infty}E\left(t-t_{\textrm{out}}\right)+\int_{t_{\textrm{out}}}^{\infty}\left\{-\frac{Z}{r\left(t\right)}\right\}. 
\label{phase_1}
\end{align}
The linearly divergent energy contribution $\lim_{t\to\infty}E\left(t-t_{\textrm{out}}\right)$ is to be disregarded, since it does not yield a relative phase between the trajectories leading to the same bin of momentum space. The same is also true for the finite energy contribution $E\left(t_{\textrm{out}}-\tau\right)$. The divergent Coulomb phase, $-Z\int_{t_{\textrm{out}}}^{\infty}dt/r\left(t\right)$, can be regularized, and the corresponding finite contribution is calculated analytically, see Ref.~\cite{Shvetsov2016}. In the second case, an electron leaves the circle $r=R_{0}$ before the laser pulse terminates, i.e., $t_{\textrm{out}}<\tau$. The corresponding SCTS phase is given by:
\begin{align}
\Phi\left(t_0,\vec{v}_{0}\right)&=\int_{t_0}^{t_{\textrm{out}}}dt\left\{\frac{p^2\left(t\right)}{2}+V_{\textrm{ion}}\left[\vec{r}\left(t\right)\right]-\vec{r}\left(t\right)\cdot \vec{\nabla}V_{\textrm{ion}}\left[\vec{r}\left(t\right)\right]\right\} \nonumber \\
&+\int_{t_{\textrm{out}}}^{\tau}dt\left\{\frac{p^2}{2}-\frac{2Z}{r}\right\}+\lim_{t\to \infty}E\left(t-\tau\right)+\int_{\tau}^{\infty}\left\{-\frac{Z}{r}\right\}. 
\label{phase_2}
\end{align}   
In Eq.~(\ref{phase_2}) we also disregard the divergent contribution $\lim_{t\to\infty}E\left(t-\tau\right)$ and regularize the Coulomb phase.

In the simulations we apply an importance sampling method, i.e., we distribute ionization times and initial transverse velocities in accord with the square root of the ionization probability [Eq.~(\ref{bondar_rate})]. We launch an ensemble of classical trajectories, find their final momenta, and bin the trajectories in cells in momentum space in accord with these momenta. The amplitudes $\exp\left[i\Phi\left(t_0^{j},\vec{v}_{0}^{j}\right)\right]$ $\left(j=1,...,n_{0}\right)$ of all $n_{0}$ trajectories leading to the same bin centered at a given final momentum $\vec{k}$ are added coherently, and the corresponding ionization probability for this final momentum reads as
\begin{equation}
\frac{dR}{d^3\vec{k}}=\left|\sum_{j=1}^{n_{0}}\exp\left[i\Phi\left(t_0^{j},\vec{v}_{0}^{j}\right)\right]\right|^2. 
\end{equation}
We note that it is necessary to achieve convergence with respect to both the bin size and the number of trajectories. The size of the momentum bin and the number of trajectories in the ensemble required for convergence depend on the system under study and the laser parameters. Typically we achieve converence at $4\times10^{8}$ trajectories and the bin size of $0.0025$~a.u. Convergence tests were performed for all results provided below. 

\section{Results and discussion}

\subsection{TDSE method}

\begin{figure}
\centering
\includegraphics[trim={0 0 0 0},width=.8\textwidth]{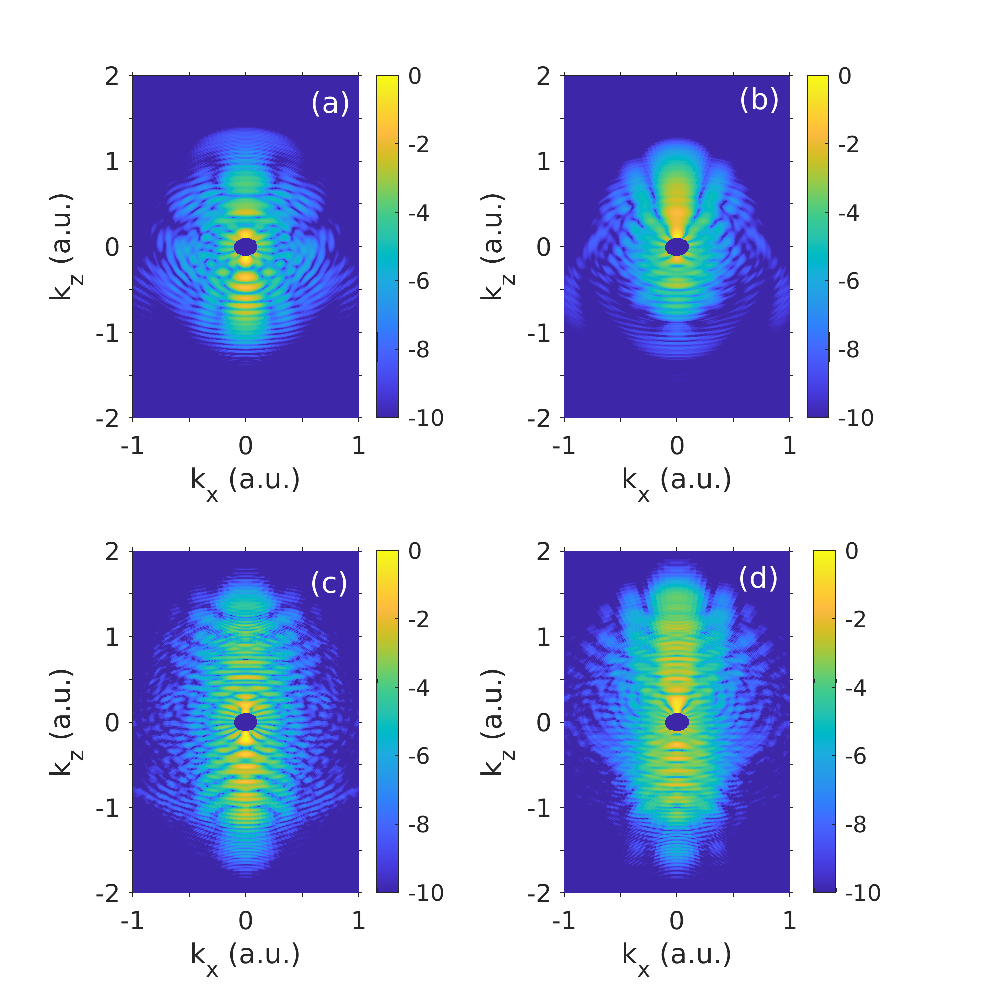}
\caption{Photoelectron momentum distributions for the CO molecule produced by $4$-cycle pulses with a wavelength of 800 nm at the intensities of (a,b) $8.8\times10^{13}$~W/cm$^2$ and (c,d) $3.51\times10^{14}$~W/cm$^2$. The PMDs in panels (a,c) show results of TDSE calculations without MEP, whereas the distributions in panels (b,d) correspond to the full account of MEP. A logarithmic color scale in arbitrary units is used. The laser field is lineraly polarized along the $z$-axis.}
\label{fig_tdse_1}
\end{figure}

In Fig.~\ref{fig_tdse_1} we show PMDs for the CO molecule probed by $4$-cycle laser pulses at intensities of $8.8\times10^{13}$ W/cm$^2$ in panels (a,b) and $3.51\times10^{14}$ W/cm$^2$ in panels (c,d). At both laser intensities, we notice the PMDs are asymmetric because of short pulse CEP effects. Nevertheless, one can see from the figure that accounting for MEP in the TDSE treatment shifts the main feature in PMDs from negative $k_z$ to positive $k_z$ momenta. Moreover, the innermost structure in the PMDs at $\left|k\right|\lesssim 0.25$~a.u. is also sensitive to the inclusion of MEP in the TDSE treatment: at low intensity in Fig.~\ref{fig_tdse_1} (a,b), the radial jets of the well-known fanlike interference structure \cite{Arbo06,Moshammer09,Arbo062,Arbo08} get less resolved upon inclusion of MEP. At high intensity in Fig.~\ref{fig_tdse_1} (c,d), both the resolution and relative intensity of those radial jets get reduced significantly upon inclusion of MEP.

\begin{figure}
\centering
\includegraphics[trim={0 0 0 0},width=.8\textwidth]{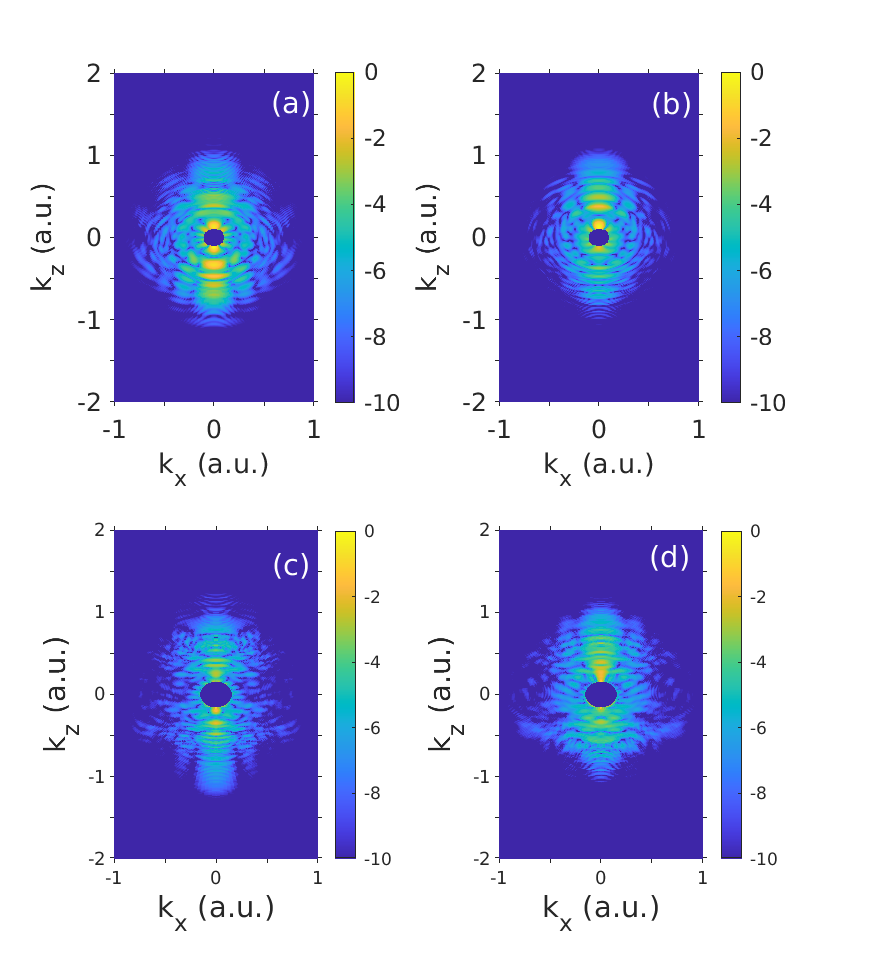}
\caption{Photoelectron momentum distributions for the CO molecule produced by $8$-cycle pulses with a wavelength of 800 nm at the intensities of (a,b) $8.8\times10^{13}$~W/cm$^2$ and (c,d) $3.51\times10^{14}$~W/cm$^2$. The PMDs in panels (a,c) show results of TDSE calculations without MEP, whereas the distributions in panels (b,d) correspond to the full account of MEP. A logarithmic color scale in arbitrary units is used. The laser field is lineraly polarized along the $z$-axis.}
\label{fig_tdse_2}
\end{figure}

Next, to shed light on effect of pulse length, we show PMDs for the CO molecule probed by 8-cycle laser pulse at intensities of $8.8\times10^{13}$ W/cm$^2$ in panels Fig.~\ref{fig_tdse_2} (a,b) and $3.51\times10^{14}$ W/cm$^2$ in panels Fig.~\ref{fig_tdse_2} (c,d). The results suggest that MEP imprint in the PMDs becomes more pronounced in the case of longer pulse. For example, the radial jets seen in Fig.~\ref{fig_tdse_2} (a) are totally washed out (no longer resolved) in Fig.~\ref{fig_tdse_2}~(b).

\subsection{Semiclassical model}

\begin{figure}
\centering
\includegraphics[trim={0 0 0 0},width=.8\textwidth]{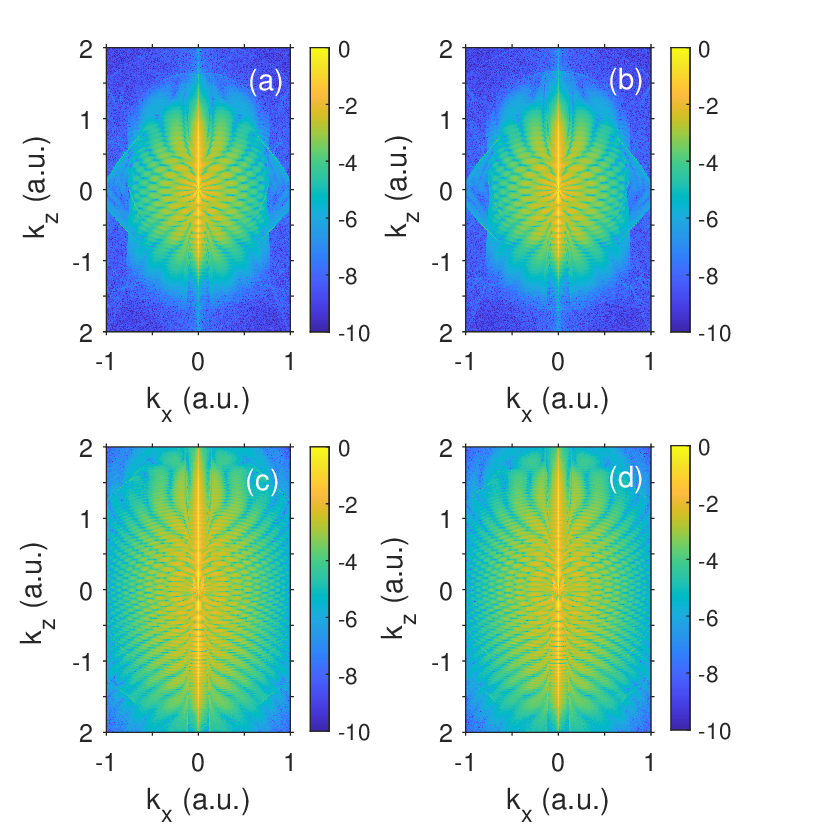}
\caption{The 2D photoelectron momentum distributions for the CO molecule ionized by a laser pulse with the intensity of $8.8\times10^{13}$~W/cm$^2$ [(a,b)] and $3.51\times10^{14}$~W/cm$^2$ [(c,d)], wavelength of $800$~nm, and the duration of 4 cycles obtained from the semiclassical model. The distributions in panels (a,c) show results of calculations disregarding MEP, whereas the PMDs in panels (b,d) are calculated accounting for MEP. A logarithmic color scale in arbitrary units is used. The laser field is linearly polarized along the $z$-axis.}
\label{fig_scts_1}
\end{figure}

\begin{figure}
\centering
\includegraphics[trim={0 0 0 0},width=.8\textwidth]{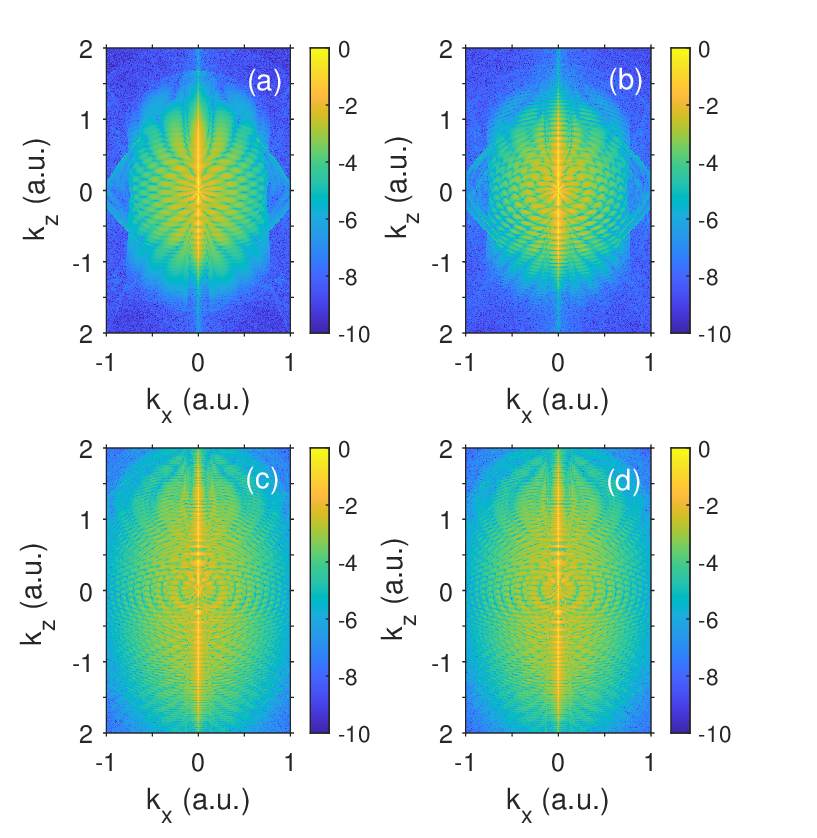}
\caption{The 2D photoelectron momentum distributions for the CO molecule ionized by a laser pulse with the intensity of $8.8\times10^{13}$~W/cm$^2$ [(a,b)] and $3.51\times10^{14}$~W/cm$^2$ [(c,d)], the wavelength of $800$~nm, and the duration of 8 cycles obtained from the semiclassical model. The distributions in panels (a,c) show results of calculations disregarding MEP, whereas the PMDs in panels (b,d) are calculated accounting for MEP. A logarithmic color scale in arbitrary units is used. The laser field is linearly polarized along the $z$-axis.}
\label{fig_scts_2}
\end{figure}

\begin{figure}
\centering
\includegraphics[trim={0 0 0 0},width=.7\textwidth]{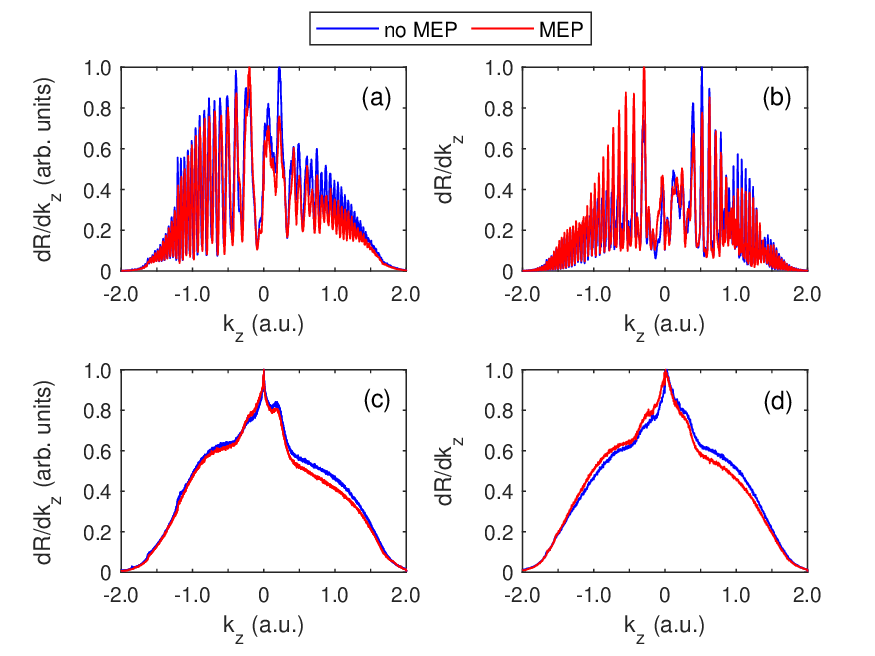}
\caption{Longitudinal momentum distributions of the photoelectrons for ionization of the CO molecule calculated from the semiclassical model with and without the MEP term. The wavelength is as in Figs.~\ref{fig_tdse_1}-\ref{fig_scts_2}, and the intensity is $3.51\times10^{14}$ W/cm$^2$. The panels [(a) and (c)] and [(b) and (d)] correspond to the pulse durations of $4$ and $8$ cycles, respectively. The panels (a,b) show the distributions calculated accounting for quantum interference, whereas the panels (c,d) display distributions obtained disregarding the interference effect. The longitudinal distributions are normalized to the peak value.}
\label{fig_long_1}
\end{figure}

The semiclassical counterparts of distributions shown in Figs.~\ref{fig_tdse_1} and \ref{fig_tdse_2} are presented in Figs.~\ref{fig_scts_1} and \ref{fig_scts_2}, respectively. It is seen from Figs.~\ref{fig_scts_1}~(a,b) and \ref{fig_scts_2} (a,b) that the MEP effect is almost invisible in semiclassical distributions for the intensity of $8.8\times10^{13}$ W/cm$^2$. Nevertheless, the distributions for the intensity of $3.51\times10^{14}$ W/cm$^2$ hold some qualitative agreements with the corresponding TDSE results as we now explain: The presence of the MEP term leads to the shift of the distribution from negative to positive longitudinal momenta $k_z$. Modifications of the innermost structure in the PMDs due to the account of MEP is also visible in Figs.~\ref{fig_scts_1}~(c,d) and \ref{fig_scts_2}~(c,d). Since at the intensity of $3.51\times10^{14}$ W/cm$^2$ the semiclassical simulations predict rich interference structures, the shift of the population due to the MEP is easier to observe in the longitudinal momentum distributions, see Figs.~\ref{fig_long_1}~(a,b).

The longitudinal distributions of Figs.~\ref{fig_long_1}~(a,b) are conveniently characterized by the asymmetry parameter $\alpha=w_{L}/w_{R}$, where $w_L=\int_{k_{z}<0}dR/dk_{z}$ and $w_R=\int_{k_{z}>0}dR/dk_{z}$ are the integrated populations for $k_z<0$ and $k_z>0$, respectively. For an 8-cycle pulse with the intensity of $3.51\times10^{14}$ W/cm$^2$ the asymmetry parameter is equal to $\alpha=0.82$ in the absence of the MEP terms, whereas $\alpha=1.06$ if these terms are taken into account. For a 4-cycle laser pulse of the same intensity, $\alpha=0.92$ in case the MEP terms are neglected and $\alpha=1.06$ when including them. Therefore, the effect is more pronounced for longer pulses which is consistent with the TDSE calculations. 

\begin{figure}[h!]
\centering
\includegraphics[trim={0 0 0 0},width=.95\textwidth]{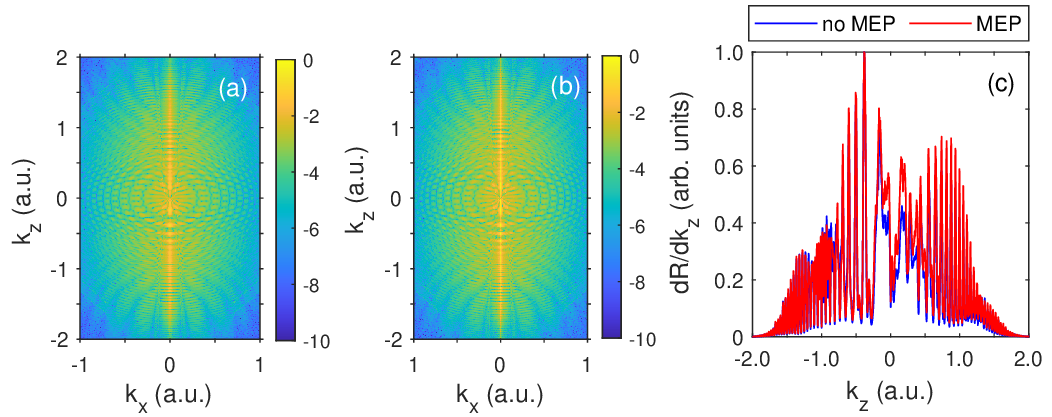}
\caption{(a,b) The two-dimensional electron momentum distributions for ionization of the CO molecule obtained using the semiclassical model neglecting the Stark-shift in Eq.~(\ref{ion_pot}). Panels (a) and (b) show distributions calculated without and with the MEP term, respectively. Panel (c) displays the corresponding longitudinal momentum distributions. The laser pulse duration is $8$ cycles, the wavelength is $800$ nm, and the intensity is $3.51\times10^{14}$~W/cm$^2$. The two-dimensional and longitudinal momentum distributions are normalized to the peak value. The laser field is linearly polarized along the $z$ axis.}
\label{fig_p_p_l}
\end{figure}

In order to understand whether the discussed MEP effect has an interference origin, we calculated the 2D PMDs and the corresponding longitudinal momentum distributions disregarding quantum interference [see Fig.~\ref{fig_long_1}~(c,d)]. To this end, we distributed ionization times and initial transverse velocities in accord with the probability distribution of Eq.~(\ref{bondar_rate}), and we calculated the number of trajectories ending up in every bin of the momentum space. Although the change of the symmetry of longitudinal distributions is less pronounced if the interference is disregarded, it is clearly visible in Figs.~\ref{fig_long_1}~(c,d). For example, at the intensity of $3.51\times10^{14}$ W/cm$^2$ and for an $8$-cycle pulse, $\alpha=0.95$ in case the MEP term is neglected, and $\alpha=1.05$ if this term is accounted for. Thus, we can conclude that the MEP effect has a kinematic origin. Further analysis of this effect shows that it depends significantly on the presence of the Stark shift. If the Stark shift is neglected, i.e., $I_{p}\left(E\right)$ is set to be equal to $I_{p}\left(0\right)$, 2D PMDs, and, as the result, the longitudinal distribution calculated with and without the MEP term are more similar to each other as compared to the case the Stark shift is accounted for, see  Figs.~\ref{fig_p_p_l}~(a-c).

\subsection{Mechanism of the MEP effect}

\begin{figure}[h!]
\centering
\includegraphics[trim={10 0 20 0},width=.95\textwidth]{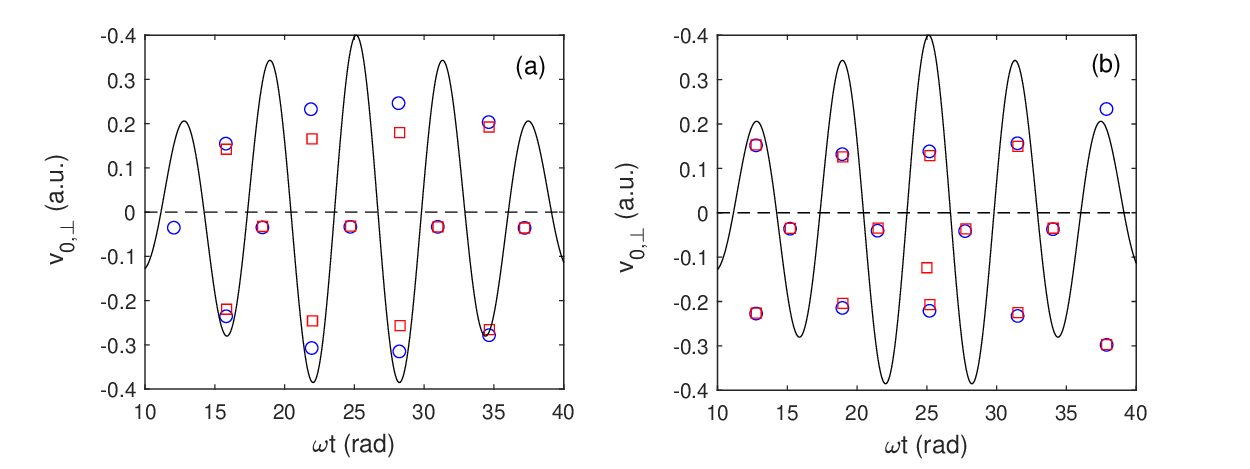}
\caption{Initial conditions (ionization time and initial transverse velocity) for the trajectories leading to the final momentum $\vec{k}=\left(-0.5,-0.05\right)$~a.u. [panel (a)] and $\vec{k}=\left(0.5,-0.05\right)$~a.u. [panel (b)]. The blue circles and red squares correspond to the trajectories calculated with and without the MEP term, respectively. The solid curve shows the laser field (in arbitrary units) given by Eq.~(\ref{laser_field}) 
The pulse parameters are as in Fig.~\ref{fig_p_p_l}.}
\label{fig_mech}
\end{figure}

To identify the mechanism responsible for the effect observed in the PMDs, we take advantage of the semiclassical model that allows us to analyze the trajectories leading to a given final momentum. We consider two final momenta symmetric with respect to $k_z=0$: $\vec{k}=\left(-0.5, -0.05\right)$~a.u. and $\vec{k}=\left(0.5,-0.05\right)$, and we find trajectories leading to the corresponding bins of momentum space. By using a hierarchical clustering algorithm (see, e.g., Ref.~\cite{EverittBook}), we group these trajectories in accord to their ionization times and initial transverse velocities and consider one representative of every group. We perform this procedure in the case where the MEP term is disregarded, and in the case where it is included, see Figs.~\ref{fig_mech}~(a) and (b). It is seen that there are two large groups of trajectories leading to the given final momenta: trajectories with relatively large values of $v_{0,\perp}$ and trajectories with small initial perpendicular velocities. Figure~\ref{fig_mech}~(a) shows that if the MEP term is accounted for, trajectories of the first group leading to $\vec{k}=(-0.5,-0.05)$~a.u. have smaller absolute values of initial transverse velocities as compared to the case where the MEP term is disregarded. Therefore, for $\vec{k}=(-0.5, -0.05)$~a.u. allowing for the MEP term leads to the shift of initial transverse velocities of the trajectories from the first group towards $\left|v_{0,\perp}\right|=0$. It is clear that trajectories with smaller values of $v_{0,\perp}$ are more probable [see Eq.~(\ref{bondar_rate})], and therefore, more trajectories lead to $\vec{k}=\left(-0.5, -0.05\right)$~a.u. if the MEP term is accounted for. Since the shift towards $\left|v_{0,\perp}\right|=0$ is absent for trajectories leading to $\vec{k}=\left(0.5, -0.05\right)$~a.u. [see Fig.~\ref{fig_mech}~(b)], electron momentum distribution restores its symmetry due to the presence of the MEP terms. 

\begin{figure}[h!]
\centering
\includegraphics[trim={10 0 20 0},width=.62\textwidth]{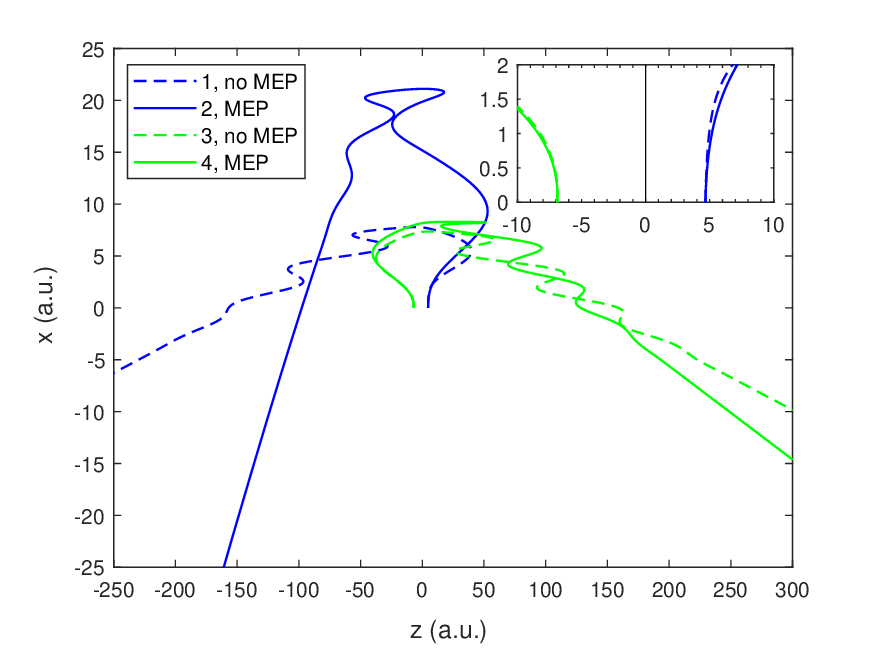}
\caption{The dashed blue (no.~1) and green (no.~3) curves show characteristic electron trajectories leading, in the absence of the MEP potential, to final momenta $\vec{k}=\left(-0.5, -0.05\right)$~a.u. and $\vec{k}=\left(0.5, -0.05\right)$~a.u., respectively. The solid blue (no.~2) and green (no.~4) curves correspond to the trajectories calculated with the account of the MEP term and starting with the same initial conditions as the trajectories no.~1 and no.~3, respectively. The inset presents a zoom-in of the initial part of the electron trajectories.}
\label{fig_mech_traj}
\end{figure}

Simultaneously, the question arises: why does the MEP term in Newton's equation of motion significantly affect trajectories leading to $\vec{k}=\left(-0.5,-0.05\right)$~a.u. and only slightly affect trajectories leading to $\vec{k}=\left(0.5,-0.05\right)$~a.u.? The reason for that is that the trajectories of the first group leading to $\vec{k}=\left(0.5,-0.05\right)$~a.u. are born close to the minima of the laser field, whereas the trajectories of same group leading to $\vec{k}=\left(-0.5,-0.05\right)$~a.u. are mostly born near the field maxima. For time instants close to the minima of the laser field, the ionization potential of Eq.~(\ref{ion_pot}), and therefore the exit point of Eq.~(\ref{sfa_exit}), are smaller than those corresponding to ionization times close to the field maxima. As a result, the MEP term has a stronger effect on trajectories leading $\vec{k}=\left(-0.5,-0.05\right)$~a.u. in the absence of this term than on trajectories that lead to $\vec{k}=\left(0.5,-0.05\right)$~a.u., see Fig.~\ref{fig_mech_traj}. Therefore, electron momentum distributions restore their symmetry due the kinematic effect caused by the Stark shift and the presence of the MEP force in equations of motion. 

\begin{figure}[h!]
\centering
\includegraphics[trim={10 0 20 0},width=.62\textwidth]{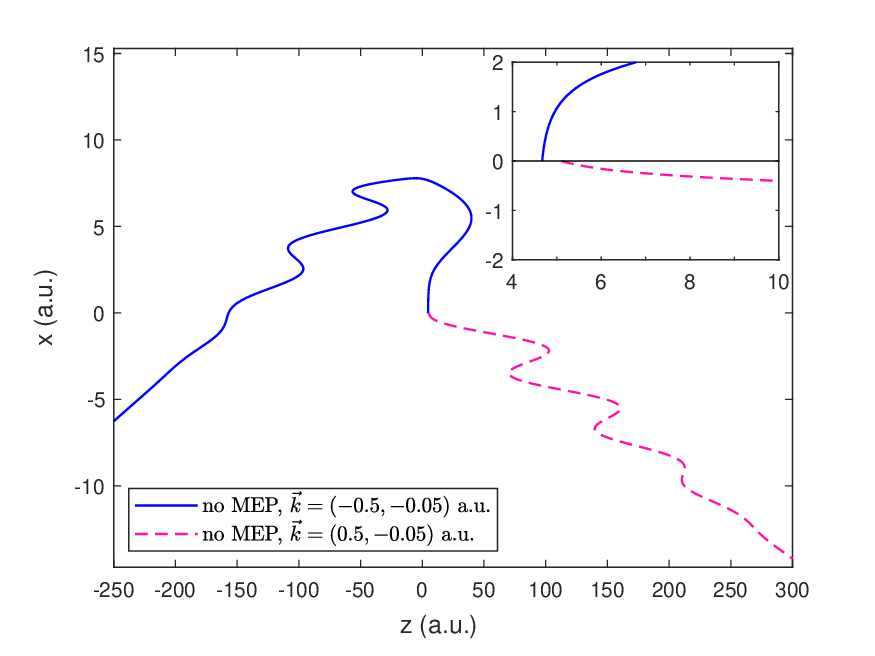}
\caption{Two characteristic electron trajectories leading, in the absence of the MEP term, to the final momenta $\vec{k}=\left(-0.5, -0.05\right)$~a.u. and $\vec{k}=\left(0.5, -0.05\right)$~a.u., respectively. The blue (solid) curve depicts the trajectory no. 1 shown in Fig.~\ref{fig_mech_traj}. The magenta (dashed) curve corresponds to the trajectory starting close to the field minima with small absolute value of initial transverse velocity. The inset presents a zoom-in of the initial part of the electron trajectories.}  
\label{fig_mech_traj2}
\end{figure}

However, the question arises regarding trajectories of the second group leading to $\vec{k}=\left(0.5,-0.05\right)$~a.u., see Fig.~\ref{fig_mech}~(b). Indeed, initial transverse velocities of these trajectories that also start close to the field minima almost do not change when the MEP term is accounted for. In order to understand the reason for this, we compare the trajectory of the first group leading to $\vec{k}=\left(-0.5,-0.05\right)$~a.u. with the trajectory of the second group that leads to $\vec{k}=\left(0.5,-0.05\right)$~a.u., see Fig.~\ref{fig_mech_traj2}. We note that the first trajectory is rescattered even though it starts with the initial transverse velocity equal to $0.23$~a.u., while the second trajectory starting with much smaller absolute value of the transverse velocity $\left|v_{0\perp}\right|=0.04$~a.u. is a direct one. Since both trajectories start near the same field maximum, their starting points are close to each other: $z_{0,1}=4.67$~a.u. and $z_{0,2}=5.10$~a.u., respectively. Simultaneously, the initial parallel velocities of these trajectories differ significantly. Since the first trajectory starts closer to the field minimum (zero of the vector potential) than the second one, its initial parallel velocity is smaller than those of the second trajectory, see Eq.~(\ref{v0_par}). Let us consider the initial parts of both trajectories adjacent to their starting points [$\left(z_{0,1},0\right)$ and $\left(z_{0,2},0\right)$, where $z_{0,1}\approx z_{0,2}$]. Since $v_{0z,2}>v_{0z,1}$, the distance from the origin at these parts of trajectories increases faster for the second trajectory than for the first one, even though $v_{0\perp,1}>v_{0\perp,2}$. As a result, the second trajectory also starting near to the field minima is much less affected by the MEP term. This explains the absence of the shift of $v_{0\perp}$ for trajectories leading to $\vec{k}=\left(0.5,-0.05\right)$~a.u. and starting close to field minima with small initial transverse velocities.

\begin{figure}[t!]
\centering
\includegraphics[trim={10 0 20 0},width=.95\textwidth]{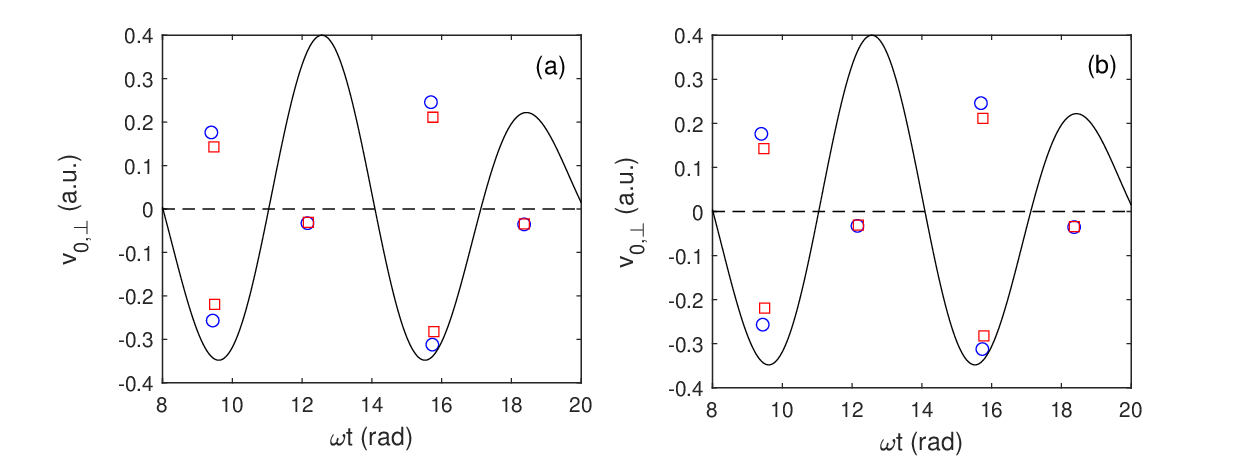}
\caption{Same as Fig.~\ref{fig_mech} but for 4-cycle laser pulse.}
\label{fig_mech_4}
\end{figure}

It remains to be explained why is the symmetry restoration effect is less pronounced in short laser pulses compared to long pulses. Figures~\ref{fig_mech_4}~(a) and (b) show ionization times and initial transverse velocities of trajectories released by the laser pulse of Eq.~(\ref{laser_field}) with a duration of $4$ optical cycles leading to $\vec{k}=\left(-0.5,-0.05\right)$~a.u. and $\vec{k}=\left(0.5,-0.05\right)$~a.u. respectively. It is seen by comparing Figs.~\ref{fig_mech} and \ref{fig_mech_4} that for the longer pulse, there are more contributing trajectories, and therefore, more of them are affected by the MEP force than for shorter pulse. Simultaneously, the shifts of the transverse initial velocities for $\vec{k}=\left(-0.5,-0.05\right)$~a.u are smaller in the shorter pulse, see Fig.~\ref{fig_mech}~(a) and Fig.~\ref{fig_mech_4}~(a). The reason for this is that for the shorter pulse with a sine-square envelope, the ionization times of trajectories from the first group correspond to smaller values of $\left|E\left(t_0\right)\right|$, and therefore, lead to larger ionization potentials than those for longer pulse. We note that for laser pulses with trapezoidal envelope this second factor responsible for the decrease of the MEP effect in shorter pulse will not work. 

\section{Conclusions and Outlook}

In conclusion, we have investigated the effect of the multielectron polarization of the parent ion on PMDs produced in strong-field ionization of the CO molecule. To this end, we have solved the TDSE and performed semiclassical simulations based on the SCTS model. In both approaches the MEP effects are described by the effective potential obtained in Refs.~\cite{Brabec2005,Zhao2007,Dimitrovski2010}. We predict a pronounced MEP effect in the 2D PMDs and longitudinal momentum distributions in the direction parallel to the laser polarization. Upon inclusion of MEP, electron momentum distributions change their asymmetry, i.e., the ratio of populations for regions with positive and negative longitudinal momenta. The effect is more pronounced for relatively long laser pulses and it weakens with decreasing pulse duration. 

By using a semiclassical model and analyzing characteristic electron trajectories we have investigated the mechanism underlying the predicted MEP effect. We have shown that the change of the asymmetry of PMDs is a kinematic effect that depends significantly on the linear Stark shift of the ionization potential and the MEP force acting on photoelectrons. More specifically, for the CO molecule the presence of the linear Stark shift leads to smaller absolute values of the tunnel exit point for trajectories starting close to the field minima as compared to the trajectories starting close to the maxima of the laser field. As a result, the trajectories launched in the vicinities of the field minima are stronger affected by the MEP force which leads to considerable changes in their final momenta. For atoms that do not have permanent dipole moment, there is no linear Stark shift. This explains the absence of the predicted MEP effects for atoms. Simultaneously, for the CO molecule we do not see a noticeable narrowing of the longitudinal momentum distributions due to the electron focusing by the induced dipole potential that was predicted for atoms, see Ref.~\cite{Shvetsov2018}. This may be attributed to the small value of the ionic polarizabilities for the CO molecule as compared to the corresponding values for Mg and Ca used in the study of Ref.~\cite{Shvetsov2018}. The effect predicted here can be experimentally verified: the necessary measurements of the asymmetry are feasible with currently existing laser systems. It is also of interest to consider this effect for other polar molecules used in strong-field ionization experiments, e.g., OCS. This will be the subject of further studies. 

\section{Acknowledgements}

This work was supported by the Deutsche Forschungsgemeinschaft (Project No.~336041027). The TDSE results presented in this work were obtained at the Centre for Scientific Computing (CSCAA), \r{A}rhus.

\end{document}